\documentclass[aps,pra, twocolumn, showpacs, superscriptaddress,preprintnumbers, amsmath, epsfig, floatfix,normalem]{revtex4-2}

\usepackage{graphicx}
\usepackage{amsmath,amssymb}
\usepackage{mathrsfs}
\usepackage[utf8]{inputenc}
\usepackage{bm}
\usepackage{physics}
\usepackage{mathtools}
\usepackage{mathptmx}
\usepackage{helvet}
\usepackage{ulem}
\usepackage{dcolumn}
\usepackage{comment}
\usepackage{ulem}
\graphicspath{ {./Figures/} }
\usepackage{textcomp}
\usepackage{graphicx,color,xcolor}

\usepackage{graphicx}
\usepackage{amsmath,amssymb}
\usepackage{color}
\usepackage[colorlinks=true,linkcolor=blue,urlcolor=blue,citecolor=blue]{hyperref}
\usepackage{mathrsfs}
\usepackage[utf8]{inputenc}

\begin{document}

\title{Universality class of a spinor Bose-Einstein condensate far from equilibrium}

\author{SeungJung Huh}
\affiliation{Department of Physics, Korea Advanced Institute of Science and Technology, Daejeon 34141, Korea }
\author{Koushik Mukherjee}
\affiliation{Department of Mathematical Physics and Nanolund, Lund University, Lund 22100, Swedens}
\affiliation{Department of Physics, Indian Institute of Technology Kharagpur, Kharagpur 721302, India}
\author{Kiryang Kwon}
\affiliation{Department of Physics, Korea Advanced Institute of Science and Technology, Daejeon 34141, Korea }
\author{Jihoon Seo}
\affiliation{Department of Physics, Korea Advanced Institute of Science and Technology, Daejeon 34141, Korea }
\author{Junhyeok Hur}
\affiliation{Department of Physics, Korea Advanced Institute of Science and Technology, Daejeon 34141, Korea }
\author{Simeon I. Mistakidis}
\affiliation{Department of Physics, Indian Institute of Technology Kharagpur 721302, India}
\affiliation{Institute for Theoretical Atomic Molecular and Optical Physics, Harvard University, Cambridge 02138, USA}
\affiliation{Department of Physics, Harvard University, Cambridge 02138, USA}
\author{H. R. Sadeghpour}
\affiliation{Institute for Theoretical Atomic Molecular and Optical Physics, Harvard University, Cambridge 02138, USA}
\affiliation{Department of Physics, Harvard University, Cambridge 02138, USA}
\author{Jae-yoon Choi}
\email{jae-yoon.choi@kaist.ac.kr}
\affiliation{Department of Physics, Korea Advanced Institute of Science and Technology, Daejeon 34141, Korea }

\date{\today}
\begin{abstract}
Scale invariance and self-similarity in physics provide a unified framework to classify phases of matter and dynamical properties near equilibrium in both classical and quantum systems. This paradigm has been further extended to isolated many-body quantum systems driven far from equilibrium, where physical observables exhibit dynamical scaling with universal scaling exponents. Universal dynamics appear in a wide range of scenarios, including cosmology, quark-gluon matter, ultracold atoms, and quantum spin magnets. However, how universal dynamics depend on the symmetry of the underlying Hamiltonian in nonequilibrium systems remain an outstanding challenge. Here, we report on the classification of universal coarsening dynamics in a quenched two-dimensional ferromagnetic spinor Bose gas. We observe spatiotemporal scaling of spin correlation functions with distinguishable scaling exponents that characterize binary and diffusive fluids. The universality class of the coarsening dynamics is determined by the symmetry of the order parameter and the dynamics of the topological defects, such as domain walls and vortices. Our results provide a categorization of the universality classes of far from equilibrium quantum dynamics based on symmetry properties of the system.
\end{abstract}

\maketitle

Critical behaviour in thermodynamic equilibrium occurs in both classical and quantum realms. Such static critical phenomena can be divided into universality classes, each class described by the same set of exponents. It was realised that two systems belonging to the same static universality class may belong to different dynamical classes~\cite{Hohenberg1977}. 
Extending the concept of universality to far from equilibrium quantum many-body systems presents a formidable challenge in physics~\cite{Polkovnikov2011,Eisert2015,Ueda2020}.
Numerous experiments in myriad platforms observed universal behaviour and spatiotemporal scaling of physical observables in late-time dynamics, where scaling exponents and scaling functions are independent of microscopic details and initial conditions. 
Celebrated examples are prethermal dynamics of unitary Bose gas~\cite{Makotyn2014,Eigen2018} and wave turbulence of atomic superfluid~\cite{Galka2022}, relaxation dynamics of spin correlations~\cite{Prufer2018} and momentum distributions~\cite{Erne2018,Glidden2021,Fontaine2022}, and emergent superdiffusive spin transport~\cite{Zu2021,Wei2022,Joshi2022}.

Recently, a  comprehensive picture of universal dynamics has emerged in isolated quantum systems~\cite{Baier2001,Berges2008,Schole2012,Schmidt2012,Berges2014,Berges2015,Orioli2015}, where quantum states driven far from equilibrium undergo critical slowing down and display self-similar time evolution associated with nonthermal fixed points. 
Universal coarsening dynamics driven by annihilation of topological defects are found in quenched multi-component Bose-Einstein condensates~\cite{Kudo2013,Hofmann2014,Williamson2016,Orioli2015,Williamson2017}, which contrast the classical theory of phase ordering kinetics in that the quantum many-body systems are not in contact with a thermal bath~\cite{Bray1994}.
Generalized hydrodynamics has been developed for integrable model~\cite{Bertini2016,Castro2016}, and superdiffusive transport in quantum magnets has been predicted~\cite{Ljubotina2017,Gopalakrishnan2019,Ljubotina2019}, described by the Kardar-Parisi-Zhang universality class~\cite{Kardar1986}. However, whether and how the far from equilibrium quantum dynamics and their universality classes depend on the symmetries of the Hamiltonian and the emergent topological textures, remains unanswered.

Here, we address these questions by studying universal coarsening dynamics in a quenched strongly ferromagnetic superfluid in two dimensions (2D). 
We demonstrate that universality can be classified by: i) the symmetry of the order parameter in the post-quench phase and ii) the merging and annihilation dynamics of the associated topological defects, such as domain walls and vortices. 
By quenching the quadratic Zeeman energy (QZE), such that a phase transition is crossed, magnetic domains of relatively small size are spontaneously generated and subsequently merge to enter the coarsening stage in the long-time evolution. 
By monitoring the spin correlation functions at various hold times, we confirm that the dynamics are self-similar regardless of varying experimental conditions. 
Specifically, when the ground state after the quench has $\mathbb{Z}_2$ (spin inversion) symmetry, the domain growth dynamics can be described by the universal scaling exponent $1/z_{\rm exp}\simeq0.58(2)$ [$1/z_{{\rm sim}}\simeq0.59(1)$] in the experiment [finite size simulations]. 
At high-momentum -- the so-called  ``Porod tail"~\cite{Bray1994} -- is also observed in the structure factor, as an imprint of the universal character of the dynamics, associated with magnetic domain formation with sharp edges.
The results show that the emergent dynamics belongs to a binary fluid universality class in the inertial hydrodynamic regime~\cite{Hohenberg1977,Kudo2013,Williamson2016}.
When the Hamiltonian exhibits SO(3) spin-rotation symmetry, the characteristics of the ensuing magnetic domain coarsening are modified.  
In the diffusive growth dynamics of domain length~\cite{Williamson2017}, the experimentally (theoretically) measured  scaling exponent is $1/z_{\rm exp}\simeq0.43(3)$ [$1/z_{{\rm sim}}\simeq0.40(1)$], which belongs to the nonthermal universality class of O$(N)$ symmetric Hamiltonians~\cite{Orioli2015,Mikheev2019}.
We identify the formation of spin vortices by matter-wave interferometry and argue that their annihilation is closely related to the observed diffusive dynamics.  
Note also that the difference in the value of the above exponents with the thermodynamic limit predictions, namely $1/z\simeq 2/3$~\cite{Kudo2013,Hofmann2014,Williamson2016} and $1/z\simeq 1/2$~\cite{Orioli2015,Williamson2017,Mikheev2019}, respectively, can be attributed to the impact of finite size effects introduced by the external trap.

\begin{figure}%
\centering
\includegraphics[width=0.5\textwidth]{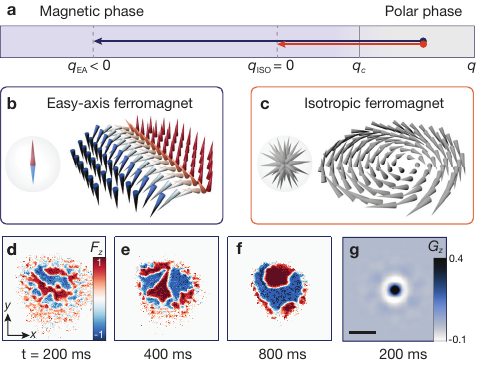}
\caption{
\textbf{Universal coarsening dynamics and topological defects.}
\textbf{a,} Schematic diagram of the experimental sequence. 
Ramping the quadratic Zeeman energy $q$, the initially prepared polar condensate is quenched to a magnetic phase $q<q_c$. 
Universal coarsening dynamics are investigated at (i) $q_{\rm EA}<0$ easy-axis ferromagnetic phases with $\mathbb{Z}_2$ spin symmetry and (ii) $q_{\rm Iso}=0$ isotropic ferromagnetic phase with SO(3) symmetry.
\textbf{b,} Cartoon picture of a magnetic domain in the easy-axis ferromagnetic phase
\textbf{c,} and a spin vortex in the isotropic ferromagnetic phase.
The magnetization vectors in each regime are shown as the spin sphere on the left side of the defects.
\textbf{d-f,} Snapshot images of magnetization at different hold times after quenching the polar condensate to easy-axis phase, $q_{\rm EA}/h=-200~\rm Hz$.
\textbf{g} Correlation function of longitudinal magnetization $G_z(x,y)$ at $t=200$~ms. 
The scale bar corresponds to 100~$\mu{}m$.
The data is averaged over 100 experimental realizations.
}
\label{fig1}
\end{figure}

\begin{figure*}[t]%
\centering
\includegraphics[width=\textwidth]{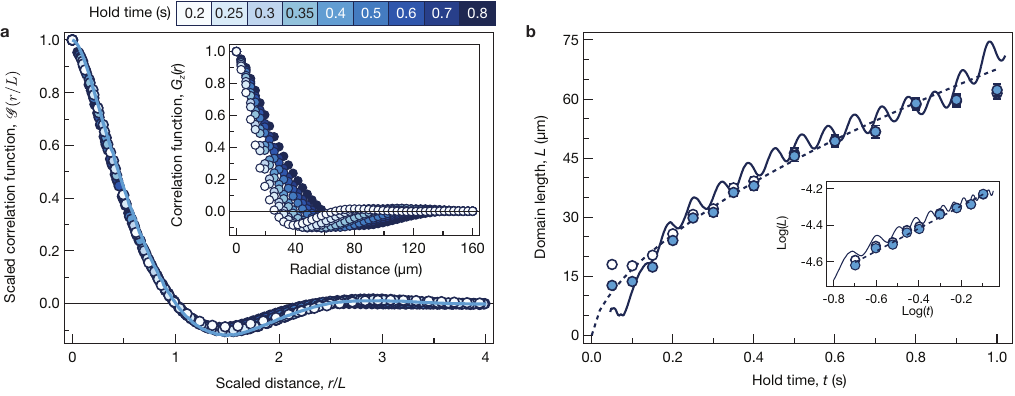}
\caption{
\textbf{Dynamic scaling and power law growth of the domain length.}
\textbf{a,} Scaled correlation function $\mathcal{G}(r/L)$ at various hold times, $t\in[0.2~\rm s, 0.8~\rm s]$. 
Longitudinal spin correlation functions at various hold times (inset) collapse onto a single function after rescaling the radial position by a domain length $L(t)$.
Here, $L(t)$ is set by a distance with $G_z(r,t)=0$. 
The solid line (light blue) represents the numerical result when using the experimental parameters.
\textbf{b,} Power law growth of the domain length $L(t)$. 
Data with closed (open) circles represent rescaled domain length after (without) deconvolution. 
The solid line is the theory line. 
The oscillatory behaviour comes from the breathing motion of the condensates.
The dashed line represents a power law function $L(t)\sim t^{1/z}$ with $1/z=0.61$, which is obtained from a linear fit in the log-log plot of the domain growth dynamics (inset).
The full dynamics of the domain length is shown in Extended Data Fig.~2, where we carefully choose the scaling range.
Small deviations observed at long evolution times between theory and experiment are attributed to atom losses by microwave dressing (Extended Data Fig.~3).
Each data point is obtained with more than 100 independent experimental runs, and one standard error of the mean (s.e.m.) is smaller than the data point. 
}
\label{fig2}
\end{figure*}

\begin{figure}[t]%
\centering
\includegraphics[width=0.46\textwidth]{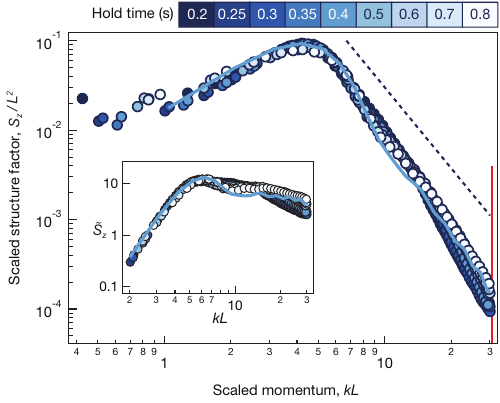}
\caption{
\textbf{Dynamic scaling of the spin structure factor in the easy-axis quench.}
The structure factor of longitudinal magnetization $S_z(k,t)$ is rescaled by the domain length: $S_z\rightarrow S_z/L^2$ and $k\rightarrow kL$.  
The dashed line is the universal Porod tail $S_z(k,t)\sim k^{-3}$ with an offset for clarity.
The vertical line (red) represents the momentum resolution for $t=0.2~\rm s$. 
As time progresses the high-momentum tail tends towards the universal $k^{-3}$ scaling. Deviations from $k^{-3}$ at higher momentum are traced back to the finite imaging resolution of the CCD pixel (3.2 $\mu$m) in the experiment and finite spatial discretization in the simulations.
Inset shows the structure factor with compensation, $\tilde{S}_z=S_z(k,t)Lk^{3}$.  
The solid line (light blue) is the numerically calculated structure factor after rescaling at $t\simeq0.8$~s.
}
\label{fig3}
\end{figure}

\section*{Ferromagnetic spin-1 system}\label{sec2}
Our experiments begin with preparing a 2D degenerate spin-1 Bose gas of $^7$Li atoms in optical dipole trap~\cite{Huh2020}. The Hamiltonian for the 2D spin-1 condensate is described by~\cite{Ho1998,Ohmi1998}.

\begin{equation}
H=\int d^2 \mathbf{r} 
\left[
\Psi^{\dagger} 
\left(
-\frac{\hbar^2\nabla^2}{2M}+qf_z^2+V_{\rm trap}
\right)
\Psi
+\frac{c_0}{2}n^2
+\frac{c_2}{2}\abs{\mathbf{F}}^2
\right],
\end{equation}
where $\Psi = ({\Psi}_1, {\Psi}_0, {\Psi}_{-1})^{T}$ is wave function of each hyperfine level ($m=-1,0,1$), $n=\Psi^{\dagger}\Psi$ is the atom density in the optical dipole trap $V_{\rm trap}$. Spin density is $\vb{F}=(F_x, F_y, F_z)$, where its $j=\{x,y,z\}$ component is $F_{j} = \Psi_{m}^{\dagger} f_{j} \Psi_{m}$ with spin-1 matrices $f_j$. 
The coefficients $c_0$ and $c_2$ represent spin-independent and spin-dependent interaction coefficients, respectively, and $q$ is the quadratic Zeeman energy. 
The $^7$Li spinor gas has ferromagnetic spin interactions ($c_2<0$), and its ground state is known to feature different phases~\cite{Kawaguchi2012} depending on the relative strength $q/\abs{c_2}n$ (Fig.1~a). For $q \gg \abs{c_2}n$, the system lies in the Polar phase where it remains unmagnetized and only the $m_z=0$ component is occupied. The region $0<q<2\abs{c_2} n$ is termed the easy-plane phase, having magnetization along the $x$-$y$ plane and all $m_z$ components being unequally populated. The $q=0$ refers to the isotropic point where all three spin components are equally populated. Entering $q < 0$, the magnetization resides in the $z$-axis and the phase is known as the easy-axis phase. 
In the experiment, the degenerate Bose gas is generated under a finite magnetic field, preparing the spinor system in the polar phase (Methods).

To initiate the nonequilibrium dynamics, we switch on the microwave field that quenches the quadratic Zeeman energy from $q/h=510$~Hz (polar phase) to a final value (Fig.~1a).  
This allows rapid cross of the phase boundaries and thus renders the initial polar state unstable, forming magnetic domains.~\cite{Sadler2006}.
After a hold time $t$, we measure the \textit{in-situ} atomic density for each spin state and record the magnetization either along the vertical $F_z$ or the horizontal spin axis $F_x$ (Methods). 
A key feature of our system is the strongly ferromagnetic spin interactions~\cite{Huh2020}, such that the characteristic time (length) scale is much shorter (smaller) compared to other alkali atomic systems. 
For instance, the spin interaction energy at the trap center is $c\simeq-h\times 160\rm~Hz$ and characteristic time scale for domain formation is $t_s=\hbar/2\abs{c}\simeq0.5~\rm{ms}$~\cite{Sadler2006}. 
Such strong interaction makes it possible to monitor the spinor gas for evolution times $t\simeq2\times10^3~t_s\simeq 1$~s, being long enough to study the emergent universal coarsening dynamics~\cite{Kudo2013,Hofmann2014,Williamson2016,Williamson2017}. 
Domain formation occurs initially at the harmonic trap center due to higher spin-dependent interaction energy. Subsequently these small magnetic domains merge and grow in the universal regime with the same power law exponent regardless of the condensate density. Therefore, we are able to investigate the universal coarsening dynamics even with the density inhomogeneity enforced by the harmonic trap.
To validate the experimental observations, we perform extensive numerical simulations of the underlying Gross-Pitaevskii equations tailored to the experimental setup.
The truncated Wigner approximation is employed~\cite{Blakie2008} accounting for quantum and thermal fluctuations in the initial polar state, see Methods for more details.

\section*{Coarsening dynamics with $\mathbb{Z}_2$ symmetry}\label{sec3}
We first investigate the nonequilibrium dynamics in the easy-axis ferromagnetic phase, $q_{\rm EA}/h=-200~\rm{Hz}$. 
The order parameter in the easy-axis has $U(1)\times\mathbb{Z}_2$ symmetry supporting the formation of magnetic domain walls as topological defects (Fig.~1b).
After the quench, the polar phase is dynamically unstable and atom pairs with $\ket{F=1,m_z=\pm1}$ ($\ket{\pm1}$) spin states and opposite momenta are generated. 
The kinetic energy, $\varepsilon_k$, of the created spin states stems from the post-quench QZE and the associated spin interaction energy, $\varepsilon_k=-q_{\rm EA}-c$~\cite{Kawaguchi2012,Kim2021}.
Since the kinetic energy is comparable to the condensate chemical potential $\mu/h=310~\rm{Hz}$, we can assure that the spinor gas is driven far from equilibrium. 
At early times, $t <10$~ms, spin-mixing takes place and the populations of the spin $\ket{\pm1}$ states increase exponentially reaching a steady value after 100~ms (Extended Data Fig.~1).
In the course of the spin-mixing process gauge vortices appear in the $\ket{\pm1}$ states,  which either annihilate or drift out of the condensate, giving their place to magnetic domains~\cite{SOM}. 
Afterward, the number of spin domains decreases and their size increases, resulting in a process known as coarsening dynamics (Fig.~1d-f).
During the coarsening dynamics, the time-evolution displays a self-similar behaviour characterized by a universal scaling law where the condensate is away from both its initial and  equilibrium state. 
For longer evolution times ($t\sim2$~s), only a few domains are left, and coarsening is terminated~({Extended Data Fig.~2}).

The scaling behaviour can be understood by analyzing the equal time correlation function of the longitudinal magnetization~\cite{Williamson2016}, $G_z(\mathbf{r},t)=\frac{1}{\mathscr{N}}\int d^2 \mathbf{r'} \langle {F_z}(\mathbf{r+r'},t){F_z}(\mathbf{r'},t) \rangle$ depicted in Fig.~1g. 
Here, $\mathbf{r}=(x,y)$ and $\mathscr{N}=\int d^2 \mathbf{r'} \langle  {F_z}(\mathbf{r'},t)^2\rangle$ is the normalization factor, which is conserved during the coarsening stage~\cite{Prufer2018}. 
In the inset of Fig.~2a, we present the radial profile of the spin correlation functions $G_z^{\rm }(r,t)$ at various hold times. 
The anti-correlation captured by $G_z^{\rm }(r,t)$ indicates the creation of magnetic domains in opposite spin states. 
We quantify, both in experiment and theory, the average domain size $L(t)$ as the first zero of the correlation function, $G_z^{\rm }(L,t)=0$~\cite{Williamson2016}. 
Indeed, upon rescaling the radial distance, $r\rightarrow r/L(t)$, the correlation functions at various hold times collapse onto a single curve, $\mathcal{G}[r/L(t)]$ (Fig.~2a), indicating the self-similar character of the universal dynamics.

The universal growth dynamics is characterized by the power law increase of the domain length $L(t)\sim t^{1/z}$ (Fig.~2b), where the dynamical critical exponent $1/z$ determines the universality class of the emergent coarsening dynamics. 
Since in the easy-axis phase, the spinor gas reduces to a binary superfluid system consisting of only the $m_{z} = \pm1$ components,  the coarsening dynamics belongs to a binary fluid universality class or Model H~\cite{Hohenberg1977}.  
Previous numerical studies operating in the thermodynamic limit indeed confirmed this argument and predicted the scaling exponent to be $1/z=2/3$~\cite{Kudo2013,Hofmann2014,Williamson2016}.

Figure~2b shows the power law growth of $L(t)$ as extracted from both experiment and theory.  
The scaling exponent in the experiment (open circles) is  $1/z_{\rm exp}=0.57(2)$, which is in excellent agreement with our mean-field simulations $1/z_{\rm sim}\simeq0.59(1)$ using the experimental parameters.  
Here, the time interval for the scaling regime is set to $t\in[0.2~\rm s, 0.8~\rm s]$, and the independency of the scaling exponent on the time interval is demonstrated (Extended Data Fig.~2a,b).
The exponents as found both experimentally and theoretically, however, are smaller than the predicted thermodynamic limit value $1/z=2/3$~\cite{Kudo2013,Hofmann2014,Williamson2016}, and we attribute this discrepancy to the finite size of our system enforced by the external trap.
While the universal scaling arguments are strictly valid in the thermodynamic limit, finite size corrections should reduce the exponent as, $1/z\simeq2/3(1-\mathcal{O}(\xi_s/L))$~\cite{Huse1986,Hofmann2014}, where $\xi_s=\hbar/\sqrt{2m\abs{c}}\simeq2.2~\mu{}m$ is the spin healing length.
This is further supported by our simulations with the harmonic trap, which gives a scaling exponent $1/z_{\rm sim}=2/3$ at large atom number ($N\simeq10^8$) ~\cite{SOM}.
Furthermore, our imaging system has an effective resolution of $5~\mu{}m$ that could increase the domain length.
Employing the Weiner deconvolution method, we recalibrate the domain size and obtain $1/z_{\rm exp}=0.61(3)$. 
Similar universal behaviour is observed in counting the magnetic domain number after the quench (Extended Data Fig.~2c).

Dynamical scaling is also represented in the structure factor, $S_z({k},t)=L(t)^2S_u({k}L(t))$, which is the fourier transformation of the spin correlation function, with a scaling function $S_u$~\cite{Williamson2016}.
The scaling form is identical to the nonthermal fixed theory that suggests $S_z(k,t)=t^{d/z}f_S(t^{1/z}k)$ in $d$ spatial dimensions, and a scaling function $f_S$~\cite{Berges2008,Schole2012,Schmidt2012,Berges2014,Berges2015,Orioli2015}. 
In Fig.~\ref{fig3}, we provide the rescaled structure factors within the time interval $t\in[0.2~\rm s, 0.8~s]$. 
A universal scaling of the Porod tail, $S_z(k)\sim k^{-3}$ is observed~\cite{Bray1994}. 
At early times ($t<100~\rm ms$), we observe that the structure factor monotonically decreases (not shown), and only after the system enters the coarsening stage the characteristic ``knee" shape and the universal high-momentum tail are revealed. 
The $k^{-3}$ scaling behavior  originates from a linear decay of the correlation function with sharp domain wall edges among the $m_z=\pm 1$ states~\cite{Bray1994}, which is confirmed in our experiment by imaging the $M_z$ and $M_x$~\cite{SOM}. 

\begin{figure}%
\centering
\includegraphics[width=0.45\textwidth]{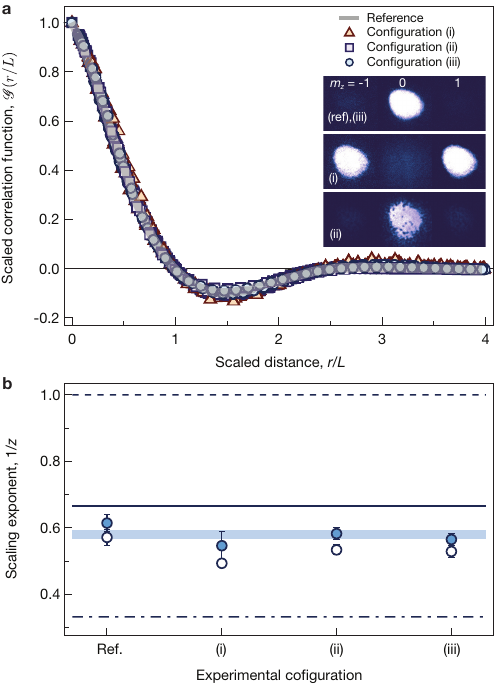}
\caption{
\textbf{Universal coarsening dynamics in the easy-axis phase.}
\textbf{a,} Scaled spin correlation functions $\mathcal{G}(r/L)$ in the coarsening stage regarding four different experimental configurations (see main text). 
The reference experiment refers to the coarsening dynamics at $q_{\rm EA}/h=-200~\rm Hz$ with a polar phase initial state (Fig.~2a).
The inset shows spin-resolved absorption images for various initial conditions.
The vortices in (ii) can be identified after 6~ms of short time-of-flight.
\textbf{b,} Dynamical scaling exponents under different configurations (Fig.~4a).
Data with closed (open) circles represent the exponent after (without) deconvolution. 
Error bars denote the fit errors (resampling error is smaller than the fit errors). 
The shaded line is the numerical calculations with finite system size, and the solid line indicates the dynamic exponents for the inertial hydrodynamic regime in the thermodynamic limit $1/z=2/3$.
The experimental results are distinguished from other dynamic exponents in a binary fluid universality class in the viscous $1/z=1$ (dashed line) and diffusive $1/z=1/3$ (dashed-dot lines) regimes.
}
\label{fig4}
\end{figure}

\begin{figure*}%
\centering
\includegraphics[width=0.9\textwidth]{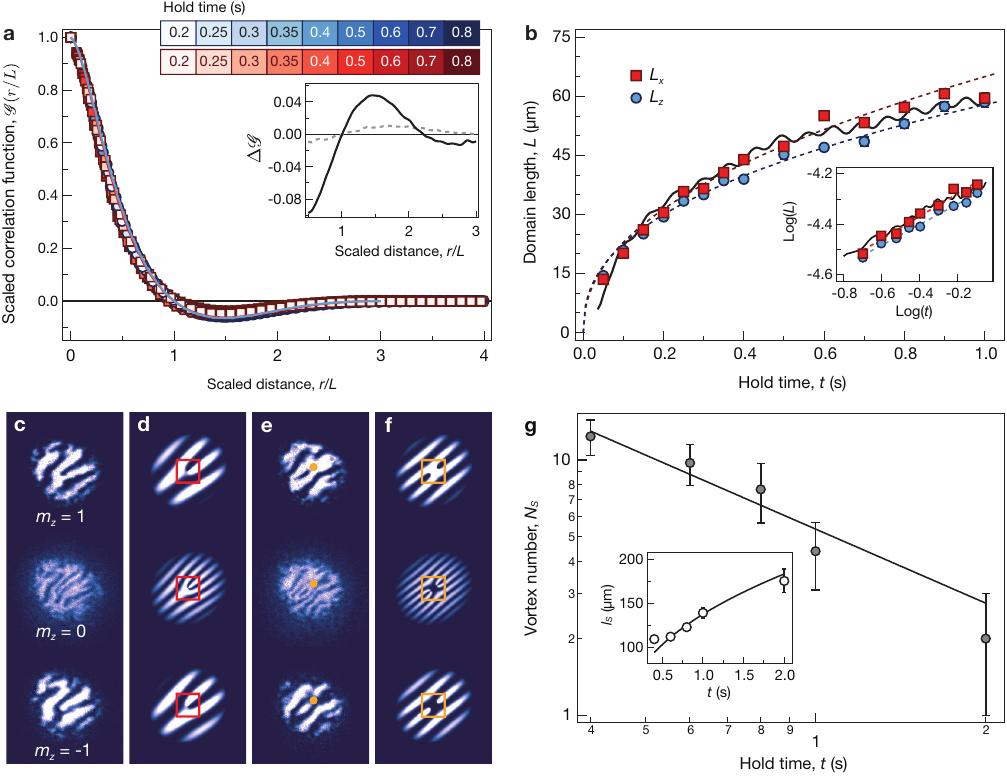}
\caption{
\textbf{Coarsening dynamics in the isotropic ferromagnetic phase.}
\textbf{a,} Scaled correlation functions of magnetization along the $x$ (red square) and $z$ axes (blue circles) from the experiment and theory (solid lines).
The domain length $L(t)$ is characterized by the first-zero of the correlation function for both axes, $G_{x,z}(L,t)=0$. 
Inset: differences in scaled correlation functions. The solid line is $\Delta \mathcal{G}=\mathcal{G}_{q_{\rm Iso}}-\mathcal{G}_{q_{\rm EA}}$, where $\mathcal{G}_{q_{\rm Iso}}=(\mathcal{G}_{x}+\mathcal{G}_{z})/2$ and $\mathcal{G}_{q_{\rm EA}}$ are obtained from interpolating data in Fig.~2a.
The dashed line is $\mathcal{G}_{x}-\mathcal{G}_{z}$. 
\textbf{b,}
Growth dynamics of the spin domain in each axis. 
The solid line is the numerical result using our experimental parameters.
The inset shows a log-log plot of the domain length dynamics, where we extract the scaling exponent by a linear fit (dashed lines). 
\textbf{c,} Matter-wave interference images during the coarsening dynamics ($t=1~\rm s$).
The two-to-one (three-to-one) fork-shape fringes in the spin $\ket{\pm1}$ ($\ket{0}$) state represent phase windings of $2\pi$ ($4\pi$) around the vortex core.
\textbf{d,} Simulated interference pattern with a spin vortex at the trap center (red boxes).
\textbf{e,} Interference images with spin vortex and anti-vortex pairs at $t=1~\rm s$ (orange circles). 
\textbf{f,} Numerical simulations with the vortex pair located at the trap center ($H$-shaped pattern, orange boxes). 
\textbf{g,} Number of spin vortices $N_{S}$ as a function of time. 
Inset: the intervortex distance $l_{S}$ during time evolution.
Solid lines are power-law guidelines, $N_{S}(t)\sim 1/L(t)^2$ and $l_{S}\sim L(t)$.
The vortex number is the average over 40 independent experiments, and the error bars indicate 1 s.e.m. }
\label{fig5}
\end{figure*}

To demonstrate universality, we further investigate the quench dynamics with three different experimental configurations (Fig.~\ref{fig4}a):
(i) We take an equal superposition of $\ket{\pm1}$ states at $q/h=510~\rm{Hz}$ as the initial state and study the quench dynamics at $q_{\rm EA}/h=-200~\rm{Hz}$. 
The initial state has different dynamical instability from the polar condensate~\cite{Kawaguchi2012,Bourges2017}, and we observe domain separation~\cite{De2014} instead of spin pair generation. 
(ii) Many vortices and anti-vortices are imprinted in the polar condensate by dragging a repulsive barrier before the quench~\cite{SOM}, and we investigate the effect of vortices on the coarsening dynamics.
(iii) We prepare the polar condensate and quench QZE to $q_{\rm EA}/h=-120~\rm{Hz}$, which is smaller than the reference experiment ($q_{\rm EA}/h=-200~\rm Hz$) but still in the easy-axis phase. 
In this case, the decay time of $L(t)$ from the microwave dressing is increased from 7~s to 40~s. 
Even with such different experimental configurations, we obtain the same universal curve upon rescaling the spin correlation function (Fig.~\ref{fig4}a), and the dynamical scaling exponents are all approximately $1/z\simeq0.58$ (Fig.~\ref{fig4}b). 
This highlights the insensitivity of the universal coarsening dynamics to experimental details, which contrasts with the near equilibrium critical phenomena~\cite{Hohenberg1977} that require a fine tuning of system parameters.
Furthermore, the scaling exponent is far different from that of other universality classes of binary fluids, such as viscous hydrodynamics with $z=1$ or diffusive dynamics with $z=3$~\cite{Bray1994}.
We reaffirm that the coarsening dynamics of the 2D ferromagnetic superfluid in the easy-axis phase belongs to the binary fluid universality class in the inertial hydrodynamic regime~\cite{Hohenberg1977}.

\section*{Coarsening dynamics with SO(3) symmetry}\label{sec4}

We now turn our attention to examining the coarsening dynamics at $q_{\rm Iso}=0$, Fig.~1c, where in contrast to the easy-axis phase the ground state is invariant under spin rotations. 
Therefore, we aim to investigate the impact of the symmetry of the order parameter, here obeying SO(3) rotational symmetry, and nature of topological defects on the universal behaviour of the spinor system. 
Since the first homotopy group of SO(3) is $\pi_1[{\rm SO(3)}]=\mathbb{Z}_2$, the condensate supports  $\mathbb{Z}_2$ spin vortices as topological defects~\cite{Kawaguchi2012}.
A recent study argued that universal coarsening dynamics also occurs at the spin isotropic point, as in the easy-axis phase, but featuring different critical exponent $1/z=0.5$~\cite{Williamson2017}.
This result is consistent with the theory of nonthermal fixed points predicting the dynamical scaling exponent $1/z\simeq 0.5$ for a bosonic Hamiltonian with O$(N)$ or U$(N)$ symmetry in $d\geq 2$ dimensions~\cite{Orioli2015,Mikheev2019}. 
Additionally, the numerical study within the mean-field framework~\cite{Williamson2017} also reported that the coarsening dynamics proceeds with the annihilation of $\mathbb{Z}_2$ spin vortices.

Figure~5 summarizes our experimental results regarding the coarsening dynamics under the spin isotropic Hamiltonian.
Since the spin vectors can point in an arbitrary direction, domain coarsening is observed in both 
{$F_x$} and {$F_z$} (Extended Data Fig.~5).
Following the same analysis as in the easy-axis quench experiment, we rescale the correlation functions by $L(t)$ and observe their collapse into a single curve (Fig.~5a), in line with the mean-field analysis. 
These universal curves are similar in each spin axis measurement but are distinctive from those of the easy-axis quench experiment (Fig.~5a inset), implying that the dynamics at the spin isotropic point belongs  to a different universality class.  
This can be further supported by the scaling exponent of the domain length $L(t)\sim t^{1/z}$ (Fig.~5b),  which we experimentally find to be $1/z_{\rm exp}\simeq0.45(3)$ for $F_x$ and $1/z_{\rm exp}\simeq0.41(2)$ for $F_z$.
Here, the exponents are close to the thermodynamic prediction $1/z=0.5$~\cite{Williamson2017} and show good agreement with our finite size numerical simulations, $1/z_{\rm sim}\simeq0.40(1)$.

To identify the underlying mechanism responsible for the coarsening dynamics  in the SO(3) phase, we monitor the spin vortices and study their decay dynamics during the coarsening stage. 
For this reason, matter-wave interferometry is adopted that can identify the position of the vortex cores by reading out the relative phase winding between spin states (Methods).
Characteristic interference patterns of the spin vortices are shown in Fig.~5c, where the fork-shaped fringes are well represented in all three spin components. 
We also observe events with closely bounded spin vortex and spin anti-vortex pairs (Fig.~5e). 
The existence of these spin vortices and vortex pairs is well reproduced in the simulated interference images (Fig.~5d, f and Extended Data Fig.~6).

Assigning the position of the spin vortex (vortex pairs) to the joint point in the fork-shaped ($H$-shaped) patterns, we count the spin vortex number $N_{S}$ and calculate the average distance between vortices $l_{S}$ at various hold times (Fig.~5g). 
The vortex number gradually decays, while the mean distance increases as time evolves. 
Since the imaging resolution is larger than the spin healing length, we underestimate the vortex number when the condensate contains many vortices.
Nevertheless, the vortex number scales with the domain size, such that $N_{S}\sim  1/L(t)^{2}$ and $l_{S}\sim L(t)$.
The decay process of spin vortex pairs occurs at a similar timescale~\cite{SOM}, hinting to its intricate connection with the universal coarsening dynamics in the isotropic SO(3) symmetric phase.
This is further supported from our numeric simulation, where we are able to calculate the argument of the transverse spin-vector and track the respective phase jumps~\cite{SOM}. \\

\section*{Conclusions and outlooks}\label{sec5}
Utilizing a strongly ferromagnetic spinor condensate, we observe universal coarsening dynamics in two dimensions. 
We find that the universal dynamics can be categorized into a well-defined universality class based on the symmetry of the order parameter and the dynamics of topological defects, such as domain walls and spin vortices. 
Our research demonstrates diverse capabilities of cold atom quantum simulators in characterizing nonequilibrium quantum dynamics, thus providing a steppingstone to a comprehensive understanding of quantum thermalization process in multidisciplinary research fields.
Further extensions include the investigation of the universal dynamics mediated by other types of excitations, such as vortices in a two-dimensional superfluid~\cite{Gauthier2019,Johnstone2019,Karl2017}, solitons in one dimension~\cite{Schmied2019a,Fujimoto2019}, magnons in the Heisenberg spin model~\cite{Bhattacharyya2020,Rodriguez2022} and chiral quantum magnetization with spin-orbit interaction~\cite{Dzyaloshinsky1958,Moriya1960}. 
Moreover, our strongly interacting platform  offers new opportunities to explore long-time thermalization dynamics in two dimensions, where the long-lived topological defects can slow down equilibration~\cite{Guzman2011}.

\section*{Acknowledgments}
We acknowledge discussions with Immanuel Bloch, Suk-Bum Chung, Fang Fang, Timon Hilker, Panayotis Kevrekidis, Kyungtae Kim, Se Kwon Kim, and Yong-il Shin. J.-y.C. is supported by the Samsung Science and Technology Foundation BA1702-06, National Research Foundation of Korea (NRF) Grant under Projects No. RS-2023-00207974 and No. 2023M3K5A1094812, and KAIST UP program.
S.I.M and H.R.S. acknowledge  support from the NSF through a grant for ITAMP at Harvard University. 
K.M. is financially supported by Knut and Alice Wallenberg Foundation (KAW No. 2018.0217) and the Swedish Research Council and also acknowledges MHRD, Govt. of India for a research fellowship at the early stages of this work.

\bibliography{Ref_Coarsening.bib}


\section*{Methods}\label{sec11}
\noindent
\textbf{Experimental systems.}
We create a spinor condensate of $^7$Li atoms in a quasi two-dimensional optical dipole trap~\cite{Huh2020}, with frequencies $(\omega_x,\omega_y,\omega_z)=2\pi\times(7,8,635)~\rm{Hz}$. 
The condensate contains $2.7\times10^6$ atoms and has negligible thermal fraction ($<5\%$). 
The chemical potential of the condensate is $\mu/h=310~\rm{Hz}$, indicating that transversal excitations are suppressed.
Indeed, no spin structures are observed along the axial direction, and thus we confirm that the coarsening dynamics occur in two dimensions.
To prepare the condensate in the polar phase, an external magnetic field of $B=1$~G is applied along the vertical axis.
Under this magnetic field, the quadratic Zeeman energy (QZE) is larger than the critical point $q>q_c=2\abs{c}$ (Fig.~1a in the main text) so that all atoms populate the same spin state $\ket{F=1,m_z=0} \equiv\ket{1,0}$.

Instantaneous quenching of QZE is experimentally realised with the aid of the microwave  dressing technique~\cite{Kim2021}. 
The QZE is given by $q = q_{\rm B} + q_{\rm MW}$, where $q_{\rm B}/h = \alpha B^2$ is the second-order Zeeman splitting of the hyperfine states of $^7$Li atoms with $\alpha=610~{\rm Hz/G^2}$, and $q_{\rm MW}/h = \Omega^2/4\delta$ denotes the respective energy shift due to the microwave field. 
In the experiment, we ramp down the external bias field to $B\simeq 100~\rm mG$ in 7~ms and simultaneously tune the microwave frequency, such that 
the QZE value is always larger than the critical point ($q>q_c$) during the field ramp.
We initiate the nonequilibrium dynamics of the spinor gas by changing the microwave frequency within $1~\mu{}\rm s$ and quench the QZE to a target point. 
The magnitude of the QZE is calibrated by studying statistics of the spin population at 2~s after the quench~\cite{Guzman2011}.
The field gradient is compensated to below $30~\mu{}\rm G/cm$ so that we observe randomly oriented domain walls even a long time after the quench ($t=2$~s). 
The microwave field induces atom loss and heating during the hold time as it couples the atoms to the $F=2$ spin state (Extended Data Fig.~3). 
It becomes noticeable in the deep easy-axis regime ($q/h=-200$~Hz), where we could underestimate the domain length and the dynamic scaling exponent. 
Fortunately, the atom loss becomes noticeable only for very long times, when all domains are merged, and the thermal fraction remains mainly below 10$\%$ throughout the evolution.
Therefore, both the atom losses and heating do not have any sizable effects on the universal coarsening dynamics.

Long-term drifts in the experimental parameters are calibrated by taking reference images. 
For example, the average magnetization at $t=2$~s is monitored every 150 min (corresponding to approximately 300 measurements), and we adjust the field gradient if needed.  
The uncertainty of the QZE is mostly due to the external field noise, which is $\pm0.3~{\rm Hz}$. 
The small fluctuations in the QZE are negligible even for the spin isotropic coarsening dynamics within the considered time scales. 
The small but nonzero QZE could lead to a coarsening transition at longer evolution times, where the scaling exponents of the domain lengths follow the easy-plane or easy-axis phase dynamics~\cite{Williamson2017}. 
The field fluctuations in the experiment are $q/q_0<10^{-3}$ so that the transition occurs after $t>1~{\rm s}~ (t>2,000~t_s)$.\\

\noindent
\textbf{In situ spin-resolved Imaging.}
Atomic density distributions in the lower hyperfine spin states are recorded using the standard absorption imaging technique after selectively converting a target spin state into an upper hyperfine state.
Under the magnetic field of 100~mG, for instance, we apply a microwave that transfers the atoms from the $\ket{1,1}$ to the $ \ket{2,1}$ state. 
The atomic distribution in the $\ket{2,1}$ state is subsequently imaged by a resonant light with $\ket{F=2}\rightarrow \ket{F'=3}$ transition.
The atoms in the other spin states are measured in a similar manner. 
Namely, after taking the first image, we apply an additional microwave pulse, flipping the hyperfine spin states from $\ket{1,0}$ ($\ket{1,-1}$) to $\ket{2,1}$ ($\ket{2,-2}$), and imaging the transferred atoms. 
To avoid cross-talk between images within the $F=1$ state, we remove all atoms in the $\ket{F=2}$ state before taking subsequent images.
This imaging sequence is also used to measure the magnetization along the spin $x$ direction, $F_x$. 
Indeed, by applying a resonant radio-frequency pulse, we rotate the measurement basis from the spin $z$ axis to the $x$ axis and record the density distribution of each spin state. 
Paradigmatic images of magnetization along $x$ and $z$ following a quench to the spin isotropic point ($q_{\rm Iso}=0$) are shown in Extended Data Fig.~5.

To characterize the imaging resolution, we prepare a spin spiral structure, which displays periodic density modulation in each spin state.
This state can be created by evolving the spin vector in the horizontal plane under a finite field gradient~\cite{Hild2014}.
The modulation period $\lambda_{\rm sp}$ is determined by the gradient strength and exposure time, and the contrast is affected by our imaging system.
For example, the contrast drops to $30\%$ at $\lambda_{\rm sp}=10~\mu{}\rm m$.
By investigating the dependence of the contrast on the wavelength $\lambda_{\rm sp}$, we can estimate that the imaging resolution is $5~\mu{}\rm m$.
The imaging parameters are optimized to have a better signal-to-noise ratio and imaging resolution.
We set the imaging light intensity to $0.5~I_{\rm sat}$ and the imaging pulse to $4~\mu{}\rm s$, where $I_{\rm sat}=2.5~\rm mW/cm^2$ is the saturation intensity. \\

\noindent
\textbf{Vortex shedding in the polar condensate.}
To demonstrate the insensitivity of the dynamical exponent on the initial conditions, we imprint many vortices in the polar phase before the quench.
An initial state containing many vortices can be prepared by adopting the vortex shedding technique~\cite{Neely2011,Kwon2015}. 
Specifically, a repulsive optical barrier is imposed at the trap center and is translated by a piezo mirror mount. 
When the speed of the barrier exceeds a critical threshold, vortices and anti-vortices are nucleated in the condensate~\cite{Neely2011}.
The optical obstacle is made of focused, blue-detuned laser light of $532~\textrm{nm}$ along the $z$-direction. Its $1/e^2$ beam waist is $8~\mu \rm m$ and the obstacle height is $V_0/\mu \approx 3$. 
The sweeping distance of the barrier from the trap center is $d=60~\mu \rm m$ and its translation speed is 6~mm/s.
The vortex cores can be identified after 6~ms of time-of-flight (inset of Fig.~4), and in particular, around 15 vortices are nucleated before the quench. 
Indeed, using this initial state containing vortices we performed the quench dynamics as described above and measured the critical exponent which remained unaffected, e.g. having the value $1/z=0.58(2)$ for the easy-axis phase.\\

\noindent
\textbf{Matter-wave interference.}
To infer the spin windings of spin vortices, we employ the matter-wave interference~\cite{choi2012,Inouye2001} with the  following experimental procedure:  A field gradient of $60~\textrm{mG/cm}$ is applied to the condensate for $t_{\rm grad}=6$~ms. Next, a resonant radio-frequency pulse is employed in order to induce the $\ket{1,0}\leftrightarrow\ket{1,\pm1}$ transition. Finally, we image all spin components in the $F=1$ state using the selective spin transfer imaging technique.
In the vortex-free state, we observe stripe lines along the gradient direction as a result of the phase accumulation across the condensate.
With the phase defect, however, the stripe lines are dislocated, where we can read out the magnitude of the relative phase winding in the spin texture. 
Because of our finite imaging resolution, we set the periodicity of the stripe patterns to be $\sim 10~\mu{}m$. 
Accordingly, this becomes the minimum distance between spin vortices detected with this scheme. 
The vortex positions in the numerical simulations, as determined by analyzing the spin vector, are the same as the phase singular points that were identified using the matter-wave interference technique (Extended Data Fig.~6). \\

\noindent \textbf{Mean-field equations of the spin-1 gas.}
The dynamics of the spin-1 condensate is described by the following set of coupled 3D Gross-Pitaevskii equations of motion  

\begin{align}\label{GP_eq}
\frac{\partial {\Psi(\vb{r};t)}}{\partial t} = -\frac{  \hslash^2 \nabla^2}{2 M} + V_{\rm trap}(\vb{r}) + q f^2_{z} + c_{0} n(\vb{r};t) + c_{2} \vb{F}(\vb{r};t) \cdot  \vb{f}. 
\end{align}
The wave function of each hyperfine $m_z=-1,0,1$ level is denoted by $\Psi = ({\Psi}_1, {\Psi}_0, {\Psi}_{-1})^{T}$, the atom mass is $\rm M$ and $\vb{r}\equiv \{x,y,z\}$. 
The spin-independent non-linear term, $c_{0}n$, is characterized by the effective strength, $c_{0} = 4 \pi \hbar^2 (a_{0} + a_{2})/(3 \rm M)$, and total density, $n=\sum_{m_z} \abs{\Psi_{m_z}}^2$. 
Here, $a_0$ and $a_2$ refer to the 3D $s$-wave scattering lengths of the atoms in the scattering channels with total spin $F = 0$ and $F = 2$ respectively.  
In contrast, the spin-dependent nonlinear term $c_2 \vb{F} \cdot \vb{f}$ accounting for interactions among the hyperfine levels contains the coupling constant $c_2 = 4 \pi \hbar^2 (a_2 -a_{0})/(3\rm M)$ and spin density $\vb{F}=(F_x, F_y, F_z)$ whose $j=\{x,y,z\}$ component is $F_{j} = \Psi_{m_z}^{\dagger} f_{j} \Psi_{m_z}$ and $f_j$ denote the spin-1  matrices. 
As discussed in the main text we focus on the $\rm ^{7}Li$ condensate possessing strong  ferromagnetic interactions, i.e. $c_2 < 0$. 
The 3D external harmonic confinement $V_{\rm trap}= \frac{1}{2} M (\omega_x^2 x^2 +\omega_y^2  y^2 + \omega^2_z z^2)$ is characterized by the in-plane trap frequency $\omega_x \approx \omega_y$ and the out-of-plane one $\omega_z$ obeying the condition $\omega_z \gg \omega_x \approx \omega_y$ which restricts the atomic motion in 2D. 
Throughout, we consider the experimentally used trap frequencies $(\omega_x, \omega_y, \omega_z) = 2 \pi \times (7, 8, 635)~\rm Hz$. 
Moreover, the length and energy scales of the system are expressed in terms of the harmonic oscillator length $l_{\rm osc} = \sqrt{\hslash/M \omega_x}$ and the energy quanta $\hslash \omega$. 
For convenience of our simulations we further cast the 3D GP Eq.~(\ref{GP_eq}) into a dimensionless form by rescaling the spatial coordinates as $x' = x/l_{{\rm osc}}$, $y' = y/l_{ {\rm osc}}$ and $z' = z/l_{{\rm osc}}$, the time as $t' = \omega_x t$ and the wave function as $\Psi_{m_{z}}(x', y', z') = \sqrt{(l^3_{ {\rm osc}}/N)} \Psi_{m_{z}}(x,y,z)$, see e.g. Ref.~\cite{Mukherjee2020,Kwon2021} for further details.

Depending on the relative strength of $q/(\abs{c_2}n)$ it is possible to realize a rich phase diagram containing first and second-order phase transitions as described in Ref.~\cite{Kawaguchi2012}. 
For instance, the quantum critical point $q = q_{c} = 2\abs{c_2} n_{\rm peak}$, with $n_{\rm peak}$ being the condensate peak density around the trap centre, separates the so-called unmagnetized polar state where all atoms are in the $m_{z}=0$ state from the easy-plane ferromagnetic phase with all $m_{z}$ states being occupied. 
The latter phase occurs within the interval $0 < q < q_c $ and it is characterized by the order parameter $\vb{F_{\perp}} = F_x +i F_y$ referring to the transverse magnetization~\cite{Williamson2017} where $F_x=[\Psi^{*}_1 \Psi_{0} + \Psi^{*}_{0}(\Psi_1 + \Psi_{-1}) + \Psi^{*}_{-1}\Psi_{0}]/\sqrt{2}$ and $F_y=i [-\Psi^{*}_1 \Psi_{0} + \Psi^{*}t_{0}(\Psi_1 - \Psi_{-1}) + \Psi^{*}_{-1}\Psi_{0}]/\sqrt{2}$. On the other hand, for $q < 0$ the ground state corresponds to an easy-axis ferromagnetic phase where the magnetization lies along the $z$-axis and the relevant order parameter is the longitudinal magnetization $F_z=\abs{\Psi_{+1}}^2-\abs{\Psi_{-1}}^2$.   \\

\noindent \textbf{Initial state preparation using quantum and thermal fluctuations.}
The initial state of the spin-1 gas (the easy-axis polar state) is realized at $q = 2q_{c}$, with the critical quadratic Zeeman shift $q_{c} = 2\abs{c_2} n_{\rm peak}/h$.   
It is represented by the wave function  

\begin{align}\label{IS}
\Psi_{\rm polar} = (0, {\Psi}_0, 0)^{T}. 
\end{align}
The latter is determined by numerically solving Eq.~\eqref{GP_eq} using the split-time Crank-Nicholson method in imaginary time~\cite{Crank1947,Antoine2013}. 
In order to trigger the dynamics, we perform a sudden change of the quadratic Zeeman coefficient $q$ from its initial value $q_{i}/ h = 510~\rm Hz$ (polar state) to a final value $q_{\rm EA}/ h = -200~\rm Hz$ (easy-axis state) or $q_{\rm Iso}/ h =0$ (isotropic point).

To monitor the system's nonequilibrium dynamics, it is essential to consider the presence of quantum and thermal fluctuations on top of the initial zero temperature mean-field state [Eq.~\eqref{IS}]. 
This contribution seeds the ensuing dynamical instabilities occurring once $q$ is quenched. 
We incorporate the impact of quantum and thermal effects exploiting the truncated Wigner approximation~\cite{Blakie2008}, namely express the wave function as $\boldsymbol{\Psi}_{I} = \boldsymbol{\Psi_{\rm polar}} +\boldsymbol{\delta}$, where $\boldsymbol{\delta} = (\delta^{+1}, \delta^{0}, \delta^{-1})^{T}$ is a noise vector constructed using the Bogoliubov quasi-particle modes of the system. 
Specifically, first, we calculate the steady state solution ($\Psi_{{\rm polar}}$), which is subsequently perturbed according to the following ansatz
$\Psi_{m_{z}} = \Psi_{{\rm polar}} + \epsilon \big(U_{m_{z}} e^{i \Omega t}  + V^{*}_{m_{z}} e^{-i\Omega^{*} t} \big )$. 
Inserting this ansatz into Eq.~\eqref{GP_eq} and linearizing with respect to the small amplitude $\epsilon$ leads to the corresponding energy eigenvalue problem which is numerically solved through diagonalization allowing us to determine the underlying eigenfrequencies $\{\Omega^{j}\}$ and eigenfunctions $\{ U^{j}_{m_{z}}, V^{j}_{m_{z}}  \}$. 
For the polar steady state, where $\Psi_{\pm 1} = 0$ and $\Psi_{0} = \Psi_{0}$, the respective eigenvalue problem reads, 

\begin{align}
\begin{pmatrix}
U_{0} \\ V_{0} \\ U_{1}\\ V_{1} \\ U_{-1} \\V_{-1}
\end{pmatrix}
\begin{pmatrix}
 a & b & 0 & 0 & 0 & 0 \\ 
 -b^{*} & -a & 0 & 0 & 0 & 0 \\
 0 & 0 & c & 0 & 0 &  d \\ 
 0 & 0 & 0 & - c & -d^{*} & 0 \\
  0 & 0 & 0 & d & c & 0 \\
  0 & 0 & -d^{*} & 0 & 0 & -c \\
\end{pmatrix}
=	\Omega \begin{pmatrix}
U_{0} \\ V_{0} \\ U_{1}\\ V_{1} \\ U_{-1} \\V_{-1}.
\end{pmatrix} 
\end{align}
Here, the matrix elements are given by $a = -\frac{1}{2} \nabla^{2}_{\perp} + V - \mu + 2 c_{0} \abs{\Psi_{0}}^2$, $b = c_{0}(\Psi_{0})^2$,
$c = -\frac{1}{2} \nabla^{2}_{\perp} + V_{{\rm trap}} - \mu + q$ and
$d = c_2 (\Psi_{0})^2$ with $\mu$ being the system's chemical potential. 
Having at hand the eigenmodes $\{ U^{j}_{m_{z}}, V^{j}_{m_{z}} \}$ and eigenvalues $\Omega^{j}$, it is possible to express the noise field in the form  
$\delta_{m_{z}} = \sum_{j > 0}^{} U^{j}_{m_{z}} \beta^{j}_{m_z} + (V^{j}_{m_{z}})^{*} (\beta^{j}_{m_z})^* $, i.e. decomposing it in terms of the low-lying collective modes of the initial state which are weighted by the random valued coefficients $\beta^{j}_{m_{z}}$. 
These coefficients are generated as 

\begin{align}\label{N_1}
\beta^{j}_{m_{z} = \pm 1} = \frac{x_j + i y_j}{2}~~~\textrm{and}~~~\beta^{j}_{m_{z} = 0} = \sqrt{(\bar{n}_{j} + \frac{1}{2})}\frac{x_j + i y_j}{\sqrt{2}}, 
\end{align}
where $x_{j}$ and $y_{j}$ are random values taken from a normally distributed Gaussian distribution characterized by zero mean and unit variance, while $\bar{n}_j = (e^{\Omega^{j} \hbar/(K_{B}T)}-1)^{-1}$ denotes the mean thermal occupation of BEC at temperature $T$. 
Notice also that the constant $1/2$ factor appearing in Eq.~\eqref{N_1} stems from the vacuum noise. 
In this sense, the initial state of the system comprises of a condensate with thermal excitations in the $m_{z} = 0$ and only vacuum noise in the $m_{z} = \pm 1$ components.  
In the experiment, the initial thermal fraction detected was less than $5\%$. To take this into account, within our simulations, we use $T=35$~nK as the system temperature for which the average thermal fraction is approximately $4\%$ of the total atom number. 
We remark that the number of thermal atoms corresponds to 
$N_{T}=\int dr [\sum_{j}\{ [(U_{m_z=0}^{j})^2 + (V_{m_z=0}^{j})^2]\bar{n}_{j}+ (V_{m_z=0}^{j})^2\}]$, see also Ref.~\cite{Blakie2008} for the relation of thermal fraction and temperature of the condensate. 
It is worth mentioning that within our simulations we have also explored the effect of varying initial temperatures but ensuring that the thermal fraction remains below $6\%$, as per the experimental conditions. Interestingly, we found that such temperature variations have no discernible impact on the coarsening dynamics.

\setcounter{figure}{0}
\renewcommand{\figurename}{Extended Data}
\renewcommand{\thefigure}{Fig. \arabic{figure}}

\begin{figure*}[h]%
\centering
\includegraphics[width=0.95\textwidth]{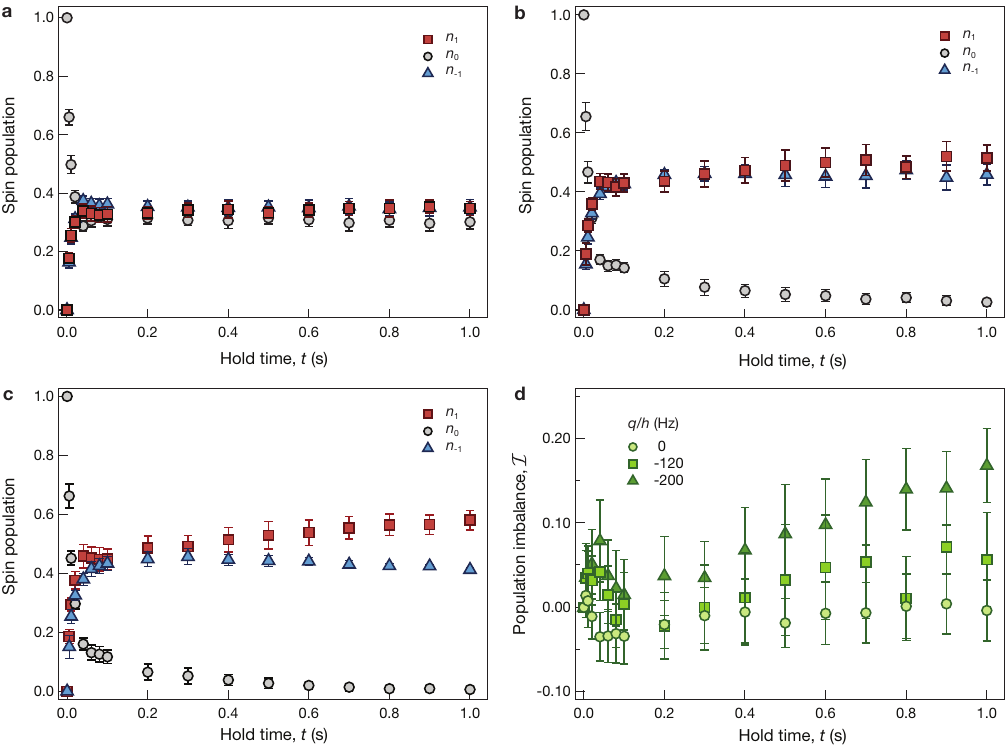}
\caption{ 
\textbf{Spin populations after quenching the quadratic Zeeman energy.}
\textbf{a,} After quenching to the isotropic ferromagnetic phase ($q/h=0$~Hz), the atoms in the spin $\ket{0}$ state rapidly decay and create spin $\ket{\pm1}$ states. 
The spin population reach a steady state after 100~ms with equal population $(n_1,n_0,n_{-1})\simeq(1/3,1/3,1/3)$.
During the whole coarsening dynamics, the spin population for all spin states remains constant.
In the easy-axis ferromagnetic phase (\textbf{b,} $q/h=-120$~Hz and \textbf{c,} $q/h=-200$~Hz), the initial $\ket{0}$ state rapidly disappear and generate equal population of the spin $\ket{\pm1}$ state. 
The residual spin component in the $\ket{0}$ state during the coarsening dynamics is attributed to the spin vector along the horizontal plane at the domain wall (Extended Data Fig.~4).
Because of the microwave dressing field, the spin population gradually changes.
\textbf{d,} Time evolution of spin population imbalance ($\mathcal{I}=n_{1}-n_{-1}$) under different quadratic Zeeman energy.
The population imbalance is noticeable in the deep easy-axis regime ($q/h=-200$~Hz), but its impact on the domain length is not significant as shown in Fig.~2. 
Each data point is obtained with more than 100 independent experimental runs, and the error bars represent one standard error of the mean.
}
\label{EFig1}
\end{figure*}

\begin{figure*}[h]%
\centering
\includegraphics[width=0.45\textwidth]{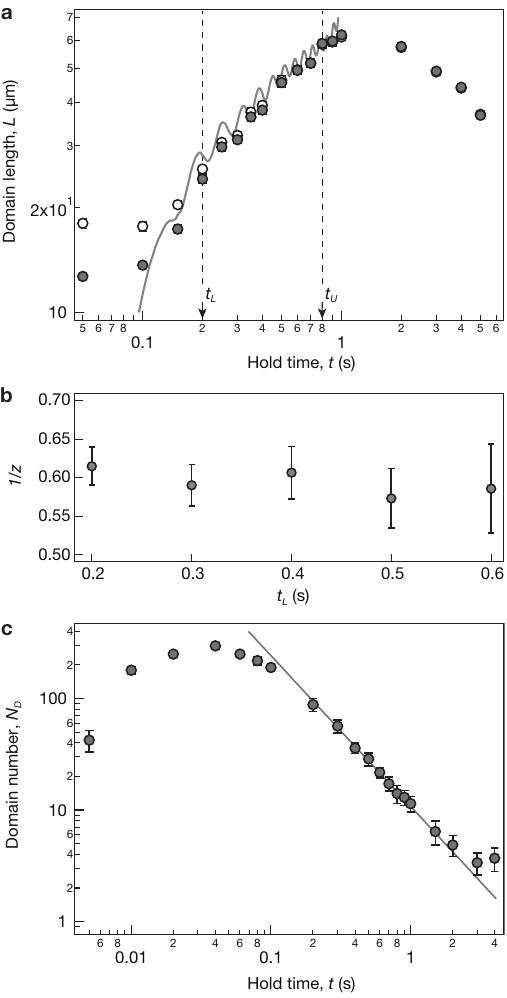}
\caption{\textbf{Full-time evolution of coarsening dynamics and scaling exponents in the easy-axis ferromagnetic phase.}
\textbf{a,} Domain length $L(t)$ in the full time evolution accessible during the experiment. Closed (open) circles represent the domain length after (without) deconvolution. Dashed lines represent the scaling time interval $t\in[0.2~\rm s, 0.8~\rm s]$. 
The lower bound for the time interval is chosen to ensure that the condensate enters into the coarsening stage after the quench.The upper bound of the time interval is limited by the finite size of the system and lifetime lifetime of the condensate.
\textbf{b,} Dependence of the scaling exponents $1/z$ on the lower bound for the time interval $t_L$. The error bars indicate the 1$\sigma$ confidence interval of the fit parameters.
\textbf{c,} Number of magnetic domains after the quench. The domain number $N_D$ is counted by using the Hoshen-Koppelman algorithm~[41]. The domain number follows the power law decay (solid line), $N_D\sim t^{-2/z}$ with $1/z=0.63(4)$.
}\label{EFig2}
\end{figure*}

\begin{figure*}[h]%
\centering
\includegraphics[width=\textwidth]{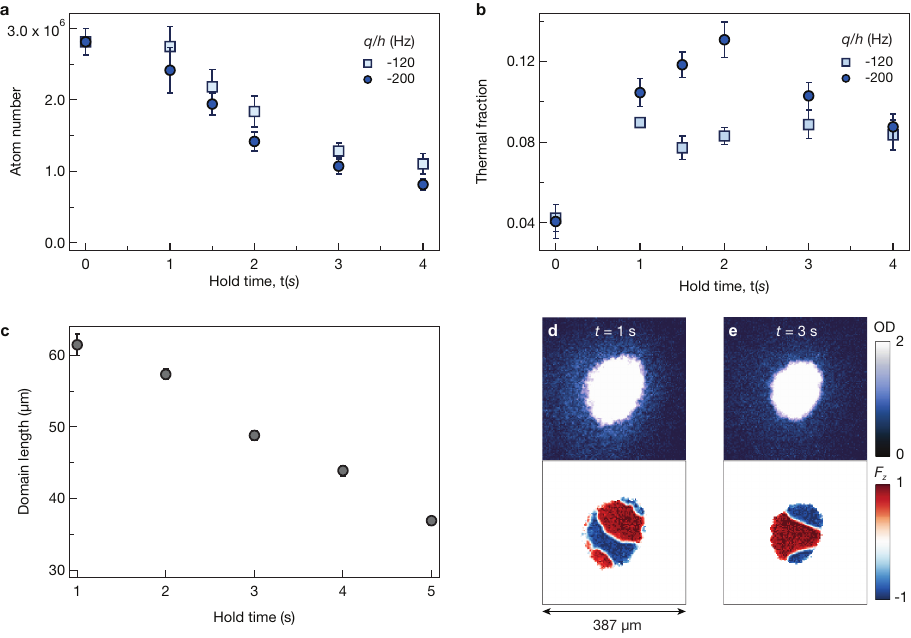}
\caption{
\textbf{Effect of the microwave dressing on the spinor Bose gas.}
\textbf{a,} Long-time evolution of the atom number and \textbf{b}, thermal fraction at $q/h=-120$~Hz (square, light blue) and $-200$~Hz (circle, dark blue). 
\textbf{c,} Long-time evolution of the domain length in the easy axis quench $q/h=-200$~Hz. After the coarsening terminated ($t=1$~s), the domain length is decreased from 61~$\mu{}$m to 57~$\mu{}$m with additional 1~s of hold time. 
It implies that the domain length during the universal dynamics $t\in[0.2 s, 0.8s] $ could be underestimated by $5\%$.  
\textbf{d,e} Absorption images (top) and the magnetization density (bottom) at different hold times (see legends).
Domain size is reduced as a result of the atom loss.
Each data point is obtained with 40 different experimental realisation, and the error bars denote one standard error of the mean.
}\label{EFig3}
\end{figure*}

\begin{figure*}[h]%
\centering
\includegraphics[width=0.75\textwidth]{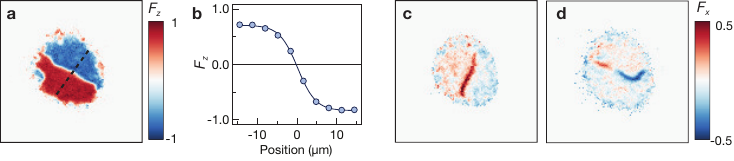}
\caption{
\textbf{Magnetic domain wall in the easy-axis ferromagnetic phase.}
\textbf{a,} Magnetization $F_z$ after 1.5~s of hold time, and \textbf{b,} the cross-section profile across the magnetic domain. 
The solid line is a fit curve $F_z(r)=F_{z0}\tanh (r/\xi_d)$ with $\xi_d=4.5(2)~\mu{}m$.
\textbf{c,d,} Magnetization along the horizontal axis $F_x$.
The spin vectors could be aligned on the same axis (c, Bloch or Neel-type domain wall) or point in opposite directions (d, Bloch line).
A long wavelength modulation of the horizontal spin vector could imply the presence of spin wave excitations in the magnetic domain.
}
\label{EFig4}
\end{figure*}

\begin{figure*}[h]
\centering
\includegraphics[width=\textwidth]{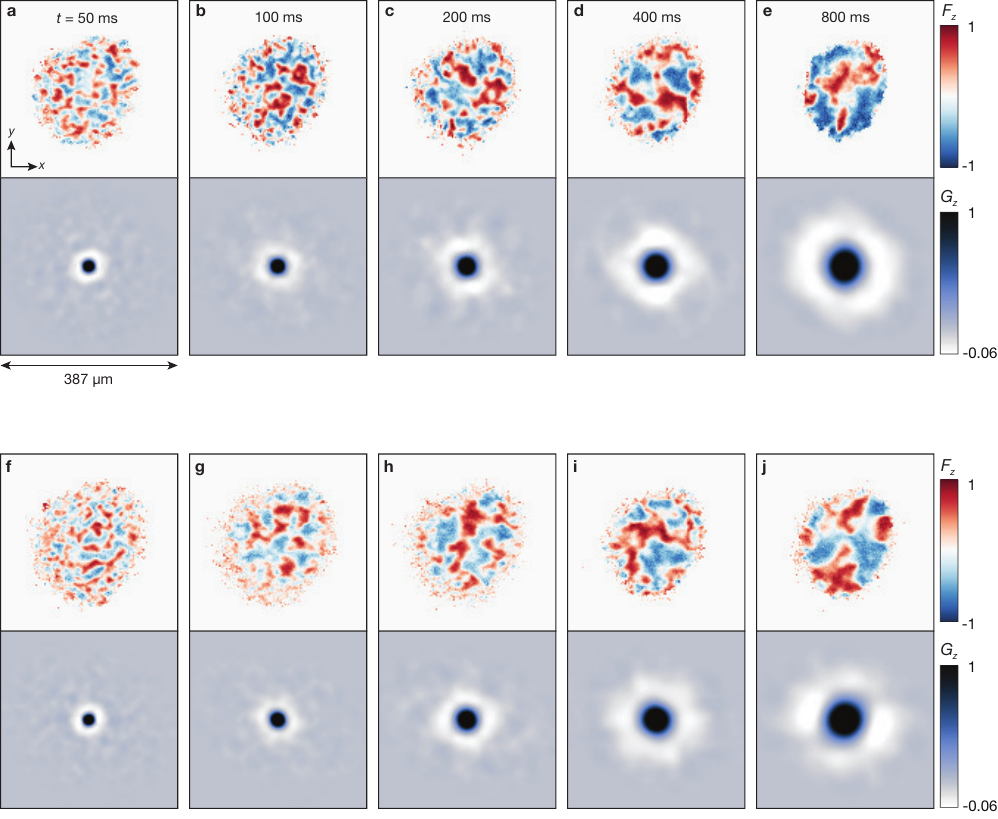}
\caption{
\textbf{Time evolution of spin domains in the isotropic ferromagnetic phase.}
\textbf{a-e,} Longitudinal magnetization $F_z$ (upper) and its two-dimensional spin correlation functions $G_z(x,y)$ (below) at various hold times. 
\textbf{f-j,} Snapshots of transverse magnetization $F_x$. and the correlation functions $G_x(x,y)$. 
In contrast to the easy-axis phase, in the spin isotropic point, coarsening dynamics are observed in both axes, and the domain boundaries are much broader than those of the $q<0$ spin domains. 
The spin correlation functions are averaged over 100 different realizations at a given hold time.
}\label{EFig5}
\end{figure*}

\begin{figure*}[t]%
\centering
\includegraphics[width=\textwidth]{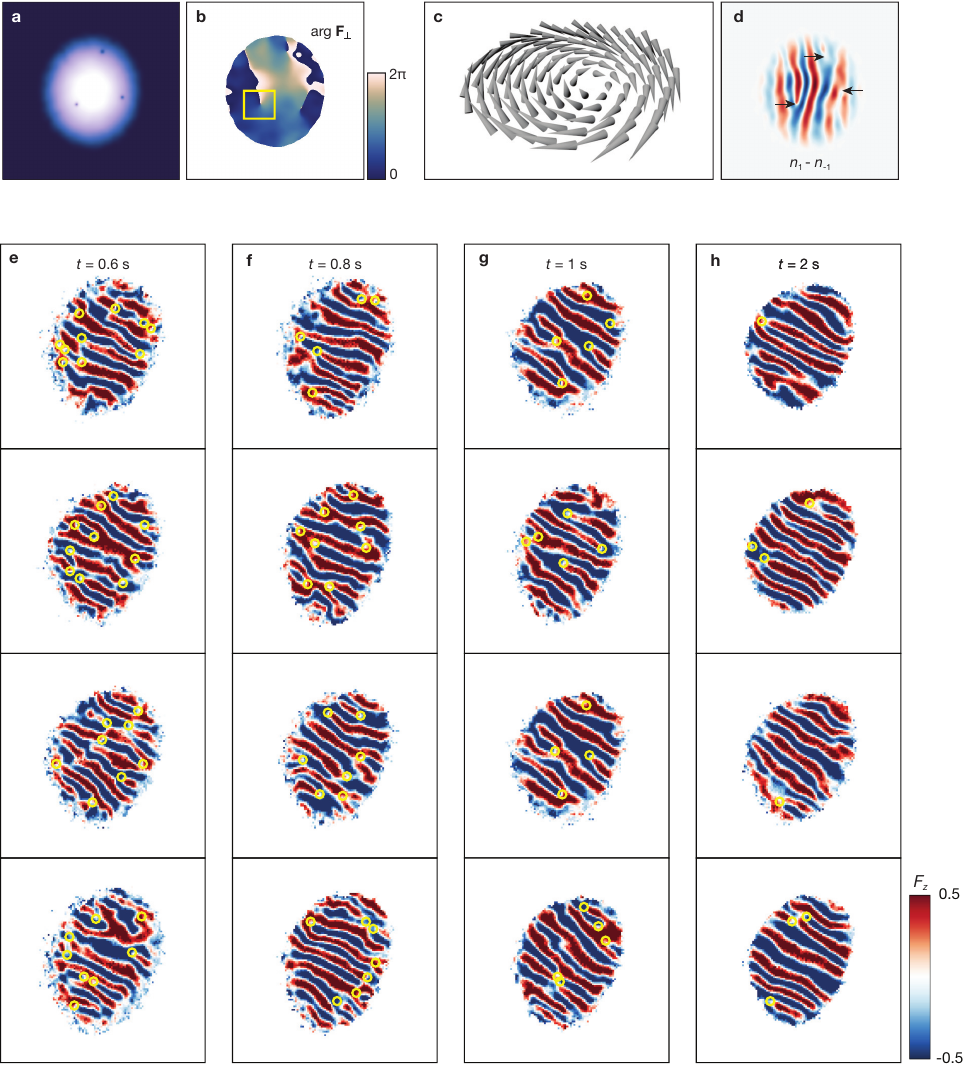}
\caption{
\textbf{Numeric simulations of the matter-wave interference for the $\mathbb{Z}_2$ spin vortices.}
\textbf{a,} Density profile and \textbf{b,} the argument of the transverse spin vector $\phi=\tan^{-1}(F_y/F_x)$ in the $x$-$y$ plane after $t=3.5~\rm s$ of domain coarsening dynamics at the spin isotropic point. 
The spin vortex can be identified from a phase jump around the vortex core (yellow box). 
\textbf{c,} Three dimensional spin vectors $\mathbf{F}=(F_x,F_y,F_z)$ in the $x$-$y$ plane near the highlighted region (yellow box). 
Not only the in-plane spin vector but also the longitudinal spin vector turns around the vortex core, indicating the $\mathbb{Z}_2$ spin vortex. 
\textbf{d,} Simulated images after the matter-wave interference. 
The positions of the spin vortices are well-identified from fork-shaped patterns in the spin imbalance image. 
Representative experimental images after the matter-wave interference under various hold times (see distinct rows).
All images are obtained by independent experimental runs.
In the vortex-free region, the magnetization displays a connected stripe pattern, while the spin vortex shows a dislocation of the stripes to form the two-to-one fork-shaped patterns. The vortex positions are highlighted by yellow circles. 
}\label{EFig6}
\end{figure*}

\end{document}